\documentclass[prc,twocolumn,showpacs,preprintnumbers,amsmath,amssymb,superscriptaddress,floatfix,nofootinbib]{revtex4}

\usepackage{graphicx}
\usepackage{amsmath}
\usepackage{amsfonts}
\usepackage{amssymb}
\usepackage{color}

\newcommand{\Slash}[1]{\ooalign{\hfil/\hfil\crcr$#1$}}

\begin{document}
\title {Regge signatures from  forward CLAS  $\Lambda(1520)$ photoproduction data}
\author{En Wang} \email{En.Wang@ific.uv.es}
\affiliation{Departamento de F\'\i sica Te\'orica and IFIC, Centro
Mixto Universidad de Valencia-CSIC, Institutos de Investigaci\'on de
Paterna, Aptd. 22085, E-46071 Valencia, Spain}
\author{Ju-Jun Xie} \email{xiejujun@impcas.ac.cn}
\affiliation{Institute of Modern Physics, Chinese Academy of
Sciences, Lanzhou 730000, China}
\affiliation{Instituto de F\'\i sica Corpuscular (IFIC), Centro
Mixto CSIC-Universidad de Valencia, Institutos de Investigaci\'on de
Paterna, Aptd. 22085, E-46071 Valencia, Spain}
\affiliation{State Key Laboratory of Theoretical
Physics, Institute of Theoretical Physics, Chinese Academy of
Sciences, Beijing 100190, China}
\author{Juan Nieves} \email{jmnieves@ific.uv.es}
\affiliation{Instituto de F\'\i sica Corpuscular (IFIC), Centro
Mixto CSIC-Universidad de Valencia, Institutos de Investigaci\'on de
Paterna, Aptd. 22085, E-46071 Valencia, Spain}
\date{\today}%

\begin{abstract}

The $\gamma p \to K^+ \Lambda(1520)$ reaction mechanism is
investigated within a Regge--effective Lagrangian hybrid approach
based on our previous study of this reaction [Physical Review C89,
015203 (2014)]. Near threshold and for large $K^+$ angles, both the
CLAS and LEPS data can be successfully described by considering the
contributions from the contact, $t$-channel $\bar K$ exchange,
$u$-channel $\Lambda(1115)$ hyperon pole, and the $s$-channel
nucleon pole and $N^*(2120)$ resonance contributions. However, for
higher energies and forward $K^+$ angles, systematic discrepancies
with data appear, which hint the possible existence of sizable
quark-gluon string mechanism effects. We show how the inclusion of a
$\bar K$ Regge--trajectory exchange in the $t$-channel leads to an
efficient description of the $\Lambda(1520)$ photoproduction channel
over the whole energy and angular ranges accessible in the CLAS
experiment.

\end{abstract}
 \maketitle

\section{Introduction}

The associate production of hadrons by photons has been extensively
studied since it provides an excellent tool to learn details of the
hadron spectrum. In particular, the $\gamma p \to K^+ \Lambda(1520)$
reaction is an efficient isospin $1/2$ filter for studying nucleon
resonances decaying to $K\Lambda(1520)$. As a consequence, the
experimental database on this reaction has expanded significantly in
recent years. In addition to the pioneering measurements at
Cornell~\cite{cornell}, CEA~\cite{CEA}, SLAC~\cite{boyarski71} and
Daresbury~\cite{Barber:1980zv} laboratories, in 2001 the CLAS
Collaboration investigated this process in
electroproduction~\cite{Barrow:2001ds} and later in 2010, this
reaction has been examined at photon energies below 2.4 GeV in the
SPring-8 LEPS experiment using a forward-angle spectrometer and
polarized photons~\cite{leps1,leps2}, and from threshold to 2.65 GeV
with the SAPHIR detector at the electron stretcher facility ELSA in
Bonn~\cite{Wieland:2011zz}. Very recently, the exclusive
$\Lambda(1520)$ photoproduction cross section has been measured by
using the CLAS detector for energies from threshold up to an
invariant $\gamma p$ mass $W = 2.85$ GeV ~\cite{Moriya:2013hwg}.

In parallel to this great experimental activity, there have also
been a large number of theoretical investigations of the
$\Lambda(1520)$ ($\equiv \Lambda^*$) resonance production with the
$\gamma p \to K^+ \Lambda(1520)$ reaction. For invariant masses $W
\leq 3$ GeV, most of these theoretical
calculations~\cite{Nam:2005uq,Nam:2006cx,Nam:2009cv,Xie:2010yk,He:2012ud,Xie:2013mua,Nam:2013nfa}
describe reasonably well the experimental data within the framework
of effective Lagrangian approach. One of the latest of these works
correspond to that of Ref.~\cite{Xie:2013mua}, where in addition to
the contact, $s$-channel nucleon pole and $t$-channel $\bar K$
exchange contributions, which were already considered in previous
works, the $s$-channel $N^*(2120)$ [previously called $N^*(2080)$]
resonance and the $u$-channel $\Lambda(1115)$ hyperon pole terms
were also included. The latter mechanism had been ignored in all
previous calculations~\cite{Xie:2010yk,He:2012ud,Nam:2005uq} that
relied on the very forward $K^+$ angular LEPS
data~\cite{leps2,leps1}, where its contribution was expected to be
small. However, it produced an enhancement for large $K^+$ angles,
and it become more and more relevant as the photon energy increases,
being essential to describe the CLAS differential cross sections at
backward angles.  On the other hand, the combined analysis of the
CLAS and LEPS data carried out in Ref.~\cite{Xie:2013mua} provided
further support on the existence of the $J^P=3/2^-$ $N^*(2120)$
resonance, and additional constraints to its properties, confirming
the previous findings of Refs.~\cite{Xie:2010yk,Nam:2013nfa}.
Indeed, the model of Ref.~\cite{Xie:2013mua} leads to an overall
good description of both sets of data, both at forward and backward
$K^+$ angles, and for the whole range of measured $\gamma p$
invariant masses in the CLAS and LEPS experiments. However, for
invariant masses $W >$ 2.35 GeV and forward angles, some small
discrepancies (though systematic) between the CLAS data and the
theoretical predictions appear (see lower panels of Fig.~3 of
Ref.~\cite{Xie:2013mua}, collected here in the right panels of
Fig.~\ref{dcs-clas}), which led to a moderate value of the best-fit
$\chi^2/dof \sim 2.5$.

This should not be entirely surprising, since the model of
Ref.~\cite{Xie:2013mua} is not suited at high energies and forward
angles, where quark-gluon string mechanisms could become
important~\cite{titovprc7274,sibiepja31, toki}. Actually, it is
obvious from the analysis of the experimental hadron cross section
data that the Reggeon and the Pomeron exchange mechanisms play a
crucial role at high energies and small transferred
momenta~\cite{Donnachie:1987pu,Grishina:2005cy}. The underlying
philosophy of the Regge formalism is as follows. In modeling the
reaction amplitude for the $\gamma p \to KY$ process at high
energies and small $|t |$ or $|u|$, instead of considering the
exchange of a finite selection of individual particles, the exchange
of entire Regge trajectories is taken into account. This exchange
can take place in the $t$ channel (kaonic trajectories) or $u$
channel (hyperonic trajectories). As such, Regge theory offers an
elegant way to circumvent the controversial issue of modeling
high-spin, high-mass particle exchange.

Different dominant mechanisms have been proposed to describe the
LAMP2 (Daresbury laboratory~\cite{Barber:1980zv}) high energy
differential cross sections. Thus, in
Refs.~\cite{titovprc7274,sibiepja31} it was claimed a large
contribution from a $t$-channel $\bar K^*$ Regge exchange. However,
in Ref.~\cite{toki}, it was argued that the $\bar K^*$ contribution
should be quite small, almost negligible, since the $K^*N\Lambda^*$
coupling is expected to be much smaller than the value implicitly
assumed in the previous works\footnote{This is because the $\Lambda
(1520)$ resonance is located very close to the threshold energy of
the $\pi \Sigma^*$ channel, which dominates the $\Lambda(1520)$
dynamics. Indeed, it could be considered as bound state of these two
hadrons, with some corrections from coupled channel dynamics. For
very small binding energies, all the couplings of the resonance tend
to zero as the mass of the bound state approaches the $\pi \Sigma^*$
threshold~\cite{Weinberg:1965zz}.}. Nevertheless, a Reggeon exchange
model, but with a $\bar K$- (instead of a $\bar K^*$) trajectory was
also used in Ref.~\cite{toki}. It was also discussed there that the
$\bar K$ Reggeon mechanism is more favored by the LAMP2 data than
the $\bar K^*$ Reggeon one, and that it is able to reproduce the
available experimental data in the region from $E^{\rm LAB}_\gamma
\sim 2.8$ GeV up to 5 GeV. Reggeized propagators for the $\bar K$
and $\bar K^*$ exchanges in the $t$-channel implemented in a
gauge-invariant manner were employed in Ref.~\cite{Nam:2010au} and
compared to Daresbury data. Note, however, that the $\bar K^*$
exchange contribution was also neglected in Ref.~\cite{Nam:2010au}.

In this work, we aim to correlate the systematic (small)
visible discrepancies, at high $\gamma p$ invariant masses and
small angles, among the theoretical predictions of
Ref.~\cite{Xie:2013mua} and the CLAS data with Regge effects.  To
this end, we improve on the model of Ref.~\cite{Xie:2013mua} by including
the contribution of a $\bar K-$Regge trajectory exchange at high energies
and low momentum transfers.  We use a hybrid model
which interpolates from the hadron effective Lagrangian approach, for
energies close to threshold, to the quark-gluon string reaction
mechanism approach, respecting gauge invariance.

Recently, it has appeared a work~\cite{He:2014gga} with similar
objectives and ideas. There, the crucial role played by the
$u$-channel $\Lambda(1115)$ hyperon pole term at backward angles is
confirmed, as well as the importance of the $N^*(2120)$ resonance to
describe the LEPS data. Moreover, Regge effects are also discussed
and taken into account, within a hybrid model that has indeed many
formal resemblances\footnote{Nevertheless, as we will explain below,
some of the parameters found in Ref.~\cite{He:2014gga} make
difficult/doubtful the theoretical interpretation of the scheme of
this reference, since $t$-channel Regge effects would have also been
considered for large scattering angles.} with the one that will be
presented in this work. However, in sharp contrast with the model
derived here, $\bar K^*$ Regge trajectory effects are considered in
Ref.~\cite{He:2014gga} and claimed to provide a considerable
contribution at high energies. It is also claimed in this reference
that the contribution from $\bar K$ and $\bar K^*$ exchange play a
similar role in the reproduction of the CLAS data. Furthermore, the
couplings of the $N^*(2120)$ state are fixed to those deduced in the
constituent quark model of Refs.~\cite{Capstick:1992uc,simonprd58},
and a large width of 330 MeV is also set for this resonance. In this
way, a great opportunity to take advantage of the accurate LEPS and
CLAS data, not only for claiming the existence of the two-star
$N^*(2120)$ state, but also for constraining/determining some of its
poorly known properties is somehow missed in the analysis carried
out in Ref.~\cite{He:2014gga}.

The present paper is organized as follows. In Sec.~\ref{sec:formalism}, we
shall discuss the formalism and the main ingredients of the model.
In Sec.~\ref{sec:results}, we will present our main results and
finally, a short summary and conclusions will be given in
Sec.~\ref{sec:conclusions}.
%
%
\section{Formalism and ingredients} \label{sec:formalism}

\subsection{Feynman amplitudes} \label{sec:feynman}

Within the effective Lagrangian approach for the
$\Lambda(1520)$ photoproduction reaction,
\begin{equation}
\gamma(k_1,\lambda) p(k_2,s_p) \to K^+ (p_1)  \Lambda^*
(p_2,s_{\Lambda^*}), \label{eq:reac}
\end{equation}
the invariant scattering amplitudes are defined as
\begin{equation}
-iT_i=\bar u_\mu(p_2,s_{\Lambda^*}) A_i^{\mu \nu} u(k_2,s_p)
\epsilon_\nu(k_1,\lambda), \label{eq:amp}
\end{equation}
where the kinematical variables ($k_1, k_2, p_1, p_2$) are defined
as in Refs.~\cite{Xie:2010yk, Xie:2013mua}, with $t$, $s$ and $u$,
the Mandelstam variables: $t=q^2_t=(k_1-p_1)^2$, $s=(k_1+k_2)^2$ and
$u=q^2_u=(p_2-k_1)^2$. On the other hand, $u_\mu$ and $u$ are
dimensionless Rarita-Schwinger and Dirac spinors, respectively,
while $\epsilon_\nu(k_1,\lambda)$ is the photon polarization vector.
In addition, $s_p$ and $s_{\Lambda^*}$ are the proton and
$\Lambda(1520)$ polarization variables, respectively. The sub-index
$i$ stands for the contact, $t$-channel antikaon exchange,
$s$-channel nucleon and $N^*(2120)$ ($\equiv N^*$) resonance pole
terms (depicted in Fig.~1 of Ref.~\cite{Xie:2010yk}) and the
$u$-channel $\Lambda$ pole mechanism (depicted in Fig.~2 of
Ref.~\cite{Xie:2013mua}). In Eq.~(\ref{eq:amp}), $A_i^{\mu\nu}$ are
the reduced tree level amplitudes which can be obtained from the
effective Lagrangian densities given in Refs.~\cite{Xie:2010yk,
Xie:2013mua}. For the sake of completeness, we also present here
these amplitudes (see Refs.~\cite{Xie:2010yk,Xie:2013mua} for some
more details):
\begin{eqnarray}
A_t^{\mu\nu} &=& -e\frac{g_{KN\Lambda^*}}{m_{K}}
\frac{1}{t-m_K^2}
q_t^\mu (q_t^\nu - p_1^\nu)\gamma_5\, f_{\rm c}, \label{eq:at} \\
A_s^{\mu\nu} &=& -e\frac{g_{KN\Lambda^*}}{m_{K}} \frac{1}{s-m_N^2}\,
p_1^\mu \gamma_5 \left[ \Slash k_1 \gamma^\nu \, f_{\rm s} +(\Slash
k_2+m_N) \gamma^\nu \, f_{\rm c} \right.  \nonumber \\
&& \left. + (\Slash k_1 +\Slash k_2+m_N) i \, \frac{\kappa_p}{2m_N}\sigma_{\nu\rho}k_1^\rho\, f_{\rm s}\right], \label{eq:as} \\
A_c^{\mu\nu} &=& e\frac{g_{KN\Lambda^*}}{m_{K}}  g^{\mu\nu}
\gamma_5\, f_{\rm c}, \label{eq:ac} \\
A_{R}^{\mu\nu} &=& \gamma_5 \left (\frac{g_1}{m_K}\Slash p_1 g^{\mu \rho}
- \frac{g_2}{m^2_K}  p_1^{\mu} p_1^{\rho}\right) \frac{\Slash k_1 + \Slash
k_2 + M_{N^*}}{s - M_{N^*}^2+iM_{N^*}\Gamma_{N^*}} \nonumber
\\ &&\times P_{\rho \sigma}
\left[\frac{ef_1}{2m_N}(k_1^{\sigma}\gamma^{\nu}-g^{\sigma\nu}\Slash
k_1) \right. \nonumber \\
&&\left. + \frac{ef_2}{(2m_N)^2} (k_1^{\sigma}k_2^{\nu}-g^{\sigma\nu}k_1 \cdot k_2) \right] f_R,
\label{eq:ar} \\
A_{u}^{\mu\nu} \!\!\!\! &=&  \!\!\!\!  \left[
\frac{h_1}{2m_{\Lambda}}(k_1^{\mu}\gamma^{\nu}- \!\!
g^{\mu\nu}\Slash k_1) + \!\! \frac{h_2}{(2m_{\Lambda})^2}
(k_1^{\mu}q_u^{\nu}- \!\! g^{\mu\nu}k_1 \cdot q_u)\right] \nonumber \\
&&  \times \frac{\Slash q_u + m_{\Lambda}}{u - m_{\Lambda}^2}
g_{KN\Lambda}
 \gamma_5 f_u. \label{eq:au}
\end{eqnarray}

Form factors, needed because the hadrons are not point-like
particles, have been also included in the above expressions. We use the
following parametrization~\cite{Haberzettl:1998eq,Davidson:2001rk}:
\begin{eqnarray}
f_i &=&\frac{\Lambda^4_i}{\Lambda^4_i+(q_i^2-M_i^2)^2},
\quad i=s,~t,~ R,~ u \\
f_c &=& f_s+f_t-f_s f_t, \quad {\rm
and} \left\{\begin{array}{l}    q_s^2=q_R^2=s,\, \cr
M_s = m_N,  \cr M_t = m_K, \cr M_R = M_{N^*}, \cr
M_u = m_{\Lambda},
\end{array}\right. \label{F1}
\end{eqnarray}
where the form of $f_c$ is chosen such that the
on-shell values of the coupling constants are reproduced
and gauge invariance is preserved.

\subsection{Regge contributions} \label{sec:regge}

We base our model on the exchange of a dominant $\bar K$ Regge
trajectory in the $t$-channel, as suggested in Ref.~\cite{toki}. The
kaon trajectory represents the exchange of a family of particles
with kaon-type internal quantum numbers. We will discuss two
different models to include the Regge contribution in the present
calculation~\footnote{We remind that when Reggeized propagators are
employed the gauge invariance is broken, and that $t$-channel Regge
effects should only be relevant for forward angles and high
energies. These two points will be addressed below.}:

\begin{itemize}
\item \underline{model A}:
In this case, the kaon Regge trajectory contribution is obtained
from the Feynman amplitude $A^{\mu\nu}_t$ of Eq.~(\ref{eq:at}) by
replacing the usual kaon pole-like Feynman propagator by a so-called
Regge propagator, while keeping the rest of the vertex structure,
i.e.,
\begin{eqnarray}
\frac{1}{t-m^2_K} \rightarrow \left( \frac{s}{s_0} \right)^{\alpha_K}
\frac{\pi\alpha^\prime_K}{ \Gamma(1+\alpha_K) {\rm sin}(\pi
\alpha_K)}\label{eq:reggeprop},
\end{eqnarray}
with $\alpha_K(t) = \alpha'_K(t-m^2_K) = 0.8~{\rm GeV}^{-2}
\times(t-m^2_K)$, the linear Reggeon trajectory associated to the
kaon quantum numbers. The constant $s_0$ is taken as the Mandelstam
variable $s$ at threshold [$s_0=(m_K + M_{\Lambda^*})^2$], and it is
introduced to fix the dimensions and to normalize the coupling
constants. This approach is similar to that followed in
Ref.~\cite{Nam:2010au}, which was also adopted in
Ref.~\cite{He:2014gga}. The scattering amplitude for the Reggeon
exchange will finally read
\begin{eqnarray}
\left( A^{\mu\nu}_t \right)^{\rm Regg} &=& -e \frac{\bar{g}_{KN\Lambda^*}}{t-m^2_K} q^\mu_t (q^\nu_t-p^\nu_1) \gamma_5 {\cal F}^{\rm Regg}_A,  \\
{\cal F}^{\rm Regg}_A (t)&=&   \left( \frac{s}{s_0} \right)^{\alpha_K}
\frac{\pi\alpha^\prime_K (t-m^2_K)}{ \Gamma(1+\alpha_K) {\rm
sin}(\pi \alpha_K)} , \label{eq:nam}
\end{eqnarray}
where $\bar{g}_{KN\Lambda^*} = g_{KN\Lambda^*} \times \hat{f}$, with
$\hat{f}$ a overall normalization factor of the Reggeon exchange
contribution. Actually, Reggeon couplings to mesons and baryons
might be, in general, different by up to a factor of
2~\cite{Grishina:2005cy}. This undetermined scale  will be fitted to
the available data.

Note that the Regge propagator of Eq.~(\ref{eq:reggeprop}) has the
property that it reduces to the Feynman propagator $1/(t- m_K^2)$ if
one approaches the first pole on the trajectory (i.e. $t\to m_K^2$,
and thus  ${\cal F}^{\rm Regg}_A\to 1$). This means that the farther
we go from the pole, the more the result of the Regge model will
differ from conventional Feynman diagram based models.

\item \underline{model B}:
In the region of negative $t$, the Reggeized propagator in
Eq.~(\ref{eq:nam}) exhibits a factorial growth\footnote{ Note,
$\left[\Gamma(1+\alpha_K) {\rm sin}(\pi \alpha_K)\right]^{-1} =
\Gamma(1-\alpha_K)/\pi\alpha_K$.}, which is in principle not
acceptable~\cite{Cassing:2001ud}. Accordingly, the authors of
Refs.~\cite{Grishina:2005cy, toki} proposed the use of a form factor
that decreased with $t$ and a simplified expression for the Regge
contribution\footnote{In Refs.~\cite{Grishina:2005cy,toki},
trajectories with a rotating ($e^{-i\pi\alpha_K(t)}$) phase, instead
of a constant phase (see for instance the discussion in
Ref.~\cite{Guidal:1997hy}) were assumed. The difference is an
additional factor $(-1)^{\alpha_K(t)}$ in Eq.~(\ref{eq:tokiRegg}),
which only affects to the interference between the Regge and
hadronic contributions. Such interference occurs only in a limited
window of $\gamma p$ invariant masses and $t$ values, that is not
well theoretically defined. Nevertheless, the CLAS data favor a
constant phase as used in Eq.~(\ref{eq:tokiRegg}).}
\begin{eqnarray}
T_{\rm Regg} \sim \frac{e \bar{g}_{KN\Lambda^*}}{m_K}
\left(\frac{s}{s_0} \right)^{\alpha_K(t)} F(t), \label{eq:tokiRegg}
\end{eqnarray}
with $F(t)$  a Gaussian form factor that accounts for the compositeness of the
external (incoming and outgoing) hadrons,
\begin{equation}
F(t) = e^{t/a^2}, \label{eq:agauss}
\end{equation}
with a typical value of the cutoff parameter $a \sim 2$ GeV. By
analogy with model A, we include in this context the Regge effects
by replacing the form factor $f_c$ in Eq.~(\ref{eq:at}) by,
\begin{eqnarray}
f_c & \to & {\hat f}\times {\cal F}^{\rm Regg}_B = {\hat f} \times \left(\frac{s}{s_0}
\right)^{\alpha_K(t)}e^{t/a^2}\, . \label{eq:toki}
\end{eqnarray}
\end{itemize}

\subsubsection{Considerations on gauge invariance}

The inclusion of Regge effects, in either of the two models
discussed above, breaks  gauge invariance. The amplitudes of the
$s$-channel $N^*(2120)$ and the $u$-channel $\Lambda(1115)$ pole
mechanisms are gauge invariant by themselves, while  some
cancelations among the $t$-channel $\bar K$ exchange, the
$s$-channel nucleon pole and contact-term contributions are needed
to fulfill gauge invariance. In the $s$-channel nucleon pole
amplitude, the terms modulated by the form factor $f_s$ are already
gauge invariant. Thus, the cancelations mentioned refer only to the
part of the $T_s$ amplitude affected by the form factor $f_c$. We
will denote this partial amplitude  as $T^*_s$. Thus, any
modification of the $t$-channel $\bar K$ exchange mechanism should
have an appropriate counterpart in the nucleon pole and contact term
contributions. To restore gauge invariance we follow the procedure
discussed in Refs.~\cite{Corthals:2005ce,Corthals:2006nz} and also
adopted in \cite{Nam:2010au}, and replace $(T^{\rm Regg}_t +  T^*_s
+  T_c)$ by
\begin{equation}
T^{\rm Regg}_t+\left( T^*_s +
  T_c \right)\times {\hat f}\times {\cal F}^{\rm Regg}_{A,B} \, .
\label{eq:gauge}
\end{equation}

\subsubsection{Hybrid hadron and Reggeon exchange model}

We propose a hybrid mechanism to study the $\gamma p \to K^+
\Lambda(1520)$ reaction in the range of laboratory photon energies
explored by the CLAS Collaboration data. At the lowest invariant
masses, near threshold, we consider the effective Lagrangian model
of Ref.~\cite{Xie:2013mua}, which amplitudes were collected in
Subsec.~\ref{sec:feynman}. However, for the higher photon energies
($W > W_0$) and at low momentum transfers ($|t| <t_0$), or
equivalently very forward $K^+$ angles, we assume that the string
quark--gluon mechanism, discussed in Subsec.~\ref{sec:regge} is
dominant. Here, $W_0$ is a certain value of the $\gamma p$ invariant
mass above which the Regge contribution starts becoming relevant.
Similar considerations apply to the Mandelstam variable $t$, and its
distinctive value $t_0$, which limits the kaon scattering angles
where the Regge behaviour is visible.  We will implement a smooth
transition/interpolation between both reaction
mechanisms~\cite{toki}, following the procedure adopted in
Ref.~\cite{Nam:2010au}. Actually, we define/parametrize this hybrid
model by using the invariant amplitudes of
Eqs.~(\ref{eq:at})--(\ref{eq:au}), but replacing the form factor
$f_c$ by $\bar{f}_c$
\begin{eqnarray}
 f_c \to \bar{f}_c & \equiv &   {\cal F}^{\,\rm Regg}_{A,B}\times
 {\hat f} \times {\cal R} + f_c(1-{\cal
 R})
\label{eq:smooth}
\end{eqnarray}
with
\begin{eqnarray}
{\cal R}&=& {\cal R}_W \times {\cal R}_t ,\\
{\cal R}_W&=& \frac{1}{1+e^{-(W-W_0)/\Delta W}}  ,\\
{\cal R}_t&=&\frac{1}{1+e^{(|t|-t_0)/\Delta t}} , \label{eq:rt}
\end{eqnarray}
where we fix $W_0=2.35$ GeV and $\Delta W=0.08$ GeV from the
qualitative comparison of the predictions of Ref.~\cite{Xie:2013mua}
with the CLAS data and from the findings of Ref.~\cite{Nam:2010au}.
In addition, we consider  $t_0$ and $\Delta t$ as free parameters
that will be fitted to data.

It is easy to understand that ${\cal R}_W$ goes to one or to zero
when $W \gg W_0$ or $W \ll W_0$, respectively, while ${\cal R}_t$
will tend to zero if $|t| \gg t_0$ and to one when $|t| \ll t_0$, as
long as $t_0$ is sufficiently bigger than $\Delta t$. In this way,
the amplitude of the reaction smoothly shifts from that determined
from Eqs.~(\ref{eq:at})--(\ref{eq:au}) for $W \ll W_0$ to another
one for $W \gg W_0$ that it is calculated using $T^{\rm Regg}_t$,
instead of $T_t$, with the replacement of Eq.~(\ref{eq:gauge})
implemented to preserve gauge invariance. Thus, Regge effects are
incorporated with the variation of ${\cal R}_W$ from zero to one.
Similar considerations apply to the variation of the Mandelstam
variable $t$. The transition from the Regge model to the the
effective Lagrangian one is controlled by the skin parameters
$\Delta W$ and $\Delta t$.

Finally, we note that gauge invariance is accomplished at any value of
${\cal R}$.

\subsection{ Differential cross section}

The unpolarized differential cross section in the center of mass (c.m.) frame for the $\gamma p \to
K^+\Lambda(1520)$ reaction reads
\begin{eqnarray}
\frac{d\sigma}{d\cos \theta_{\rm c.m.}} &=& \frac{m_N M_{\Lambda^*}|\vec{k}^{\rm \,\, c.m.}_1||\vec{p}^{\rm
\,\, c.m.}_1|}{8\pi \left(s-m^2_N \right)^2}
\sum_{\lambda,s_p,s_{\Lambda^*}} |T|^2,
\end{eqnarray}
where $\vec{k}_1^{\rm \,\, c.m.}$ and $\vec{p}_1^{\rm \,\, c.m.}$
are the photon and $K^+$ meson c.m. three-momenta, and $\theta_{\rm
c.m.}$ is the  $K^+$ polar scattering angle. The differential cross
section $d\sigma/d(\cos\theta_{\rm c.m.})$ depends on $W$ and also
on $\cos\theta_{\rm c.m.}$.

In addition to the three new free parameters ($t_0$, $\Delta t$ and
$\hat{f}$) introduced to account for Regge effects, the model of
Ref.~\cite{Xie:2013mua} already had nine free parameters: i) the
mass and width ($M_{N^*}$ and $\Gamma_{N^*}$) of the $N^*(2120)$
resonance, ii) the cut off parameters $\Lambda_s =\Lambda_t =
\Lambda_u \equiv \Lambda_B$ and $\Lambda_R$, and iii) the
$N^*(2120)$ resonance electromagnetic $\gamma NN^*$ ($ef_1$, $ef_2$)
and strong $N^*\Lambda^*K$ ($g_1$, $g_2$) couplings and the
$\Lambda(1520)$ magnetic $\gamma\Lambda\Lambda^*$ ($h_1$) one. To
reduce the number of besfit parameters, we have kept unchanged the
contribution of the $u$-channel $\Lambda$ pole contribution, and
thus we have set the $\gamma\Lambda\Lambda^*$ coupling to the value
obtained in the Fit II of Ref.~\cite{Xie:2013mua} ($h_1=0.64$). This
is justified since the contribution of the $u$-channel $\Lambda$
pole term is only important for backward $K^+$ angles, and the Regge
mechanism should only play certain role at forward angles, In
addition, we have also fixed $\Lambda_B$ to the value of 620 MeV
quoted in Ref.~\cite{Xie:2013mua}. This cutoff parameter also
appears in $T_u$, and in the definition of the form-factor $f_c$,
which following Eq.~(\ref{eq:smooth}) is replaced by $\bar f_c$ to
account for Regge effects at high energies and low momentum
transfers~\footnote{$\Lambda_B$ also appears in the definition of
the $f_s$ form-factor that affects to some pieces of the $s$-channel
nucleon pole term. These contributions are however quite small since
they are greatly suppressed by $f_s$, and do affect very little the
best fit.}.

Thus, finally, we have ten free parameters which will be fitted to
the recent differential cross section data from the
CLAS~\cite{Moriya:2013hwg} and LEPS ~\cite{leps2} experiments.

\section{Numerical results and discussion} \label{sec:results}
\begin{table*}
\caption{Values of some parameters determined in this work and in
Ref.~\cite{Xie:2013mua}. Model A(B) parameters have been adjusted to
the combined LEPS~\cite{leps2} and CLAS~\cite{Moriya:2013hwg}
$\gamma p \to K^+ \Lambda(1520)$ $d\sigma/d(\cos\theta_{\rm c.m.})$
data including Regge effects as discussed in Eq.~(\ref{eq:nam})
(Eq.~(\ref{eq:toki})). In the last column, we compile some results
from Fit II of Ref.~\cite{Xie:2013mua}, where the mechanism of
Reggeon exchange was not considered. Finally, we also give for each
fit, the predicted $N^*(2120)$ partial decay width $\Gamma_{N^* \to
\Lambda^*K}$, and the helicity amplitudes for the positive-charge
$N^*$ state.} \vspace{0.4cm}
\begin{tabular}{cccc}
\hline \hline & \multicolumn{2}{c}{This work} & Ref.~\cite{Xie:2013mua}  \\
  & model A & model B & Fit II \\ \hline
         $g_1$ & $1.3 \pm 0.2$ & $1.4 \pm 0.2$  & $1.6 \pm 0.2$ \\
         $g_2$ & $0.9 \pm 0.5$ & $1.1 \pm 0.5$  & $2.2 \pm 0.5$ \\
          $\Lambda_R$ [MeV] & $1252 \pm 78$ & $1259 \pm 76$ & $1154 \pm 47$\\
         $ef_1$ & $0.134 \pm 0.016$ & $0.123 \pm 0.015$ & $0.126 \pm
0.012$ \\
         $ef_2$  & $-0.110 \pm 0.014$ & $-0.100 \pm 0.013$ & $-0.097 \pm
0.010$\\
         $M_{N^*}$[MeV] & $2146 \pm 5$ & $2145 \pm 5$  & $2135 \pm 4$\\
         $\Gamma_{N^*}$[MeV] & $174 \pm 14$ & $171 \pm 13$ & $184 \pm 11$\\
         $t_0$[GeV$^2$] & $0.73 \pm 0.04$ & $0.94\pm 0.05$ & $-$ \\
         $\Delta t$[GeV$^2$] & $0.28 \pm 0.02$ & $0.30 \pm 0.04$  & $-$ \\
         $\hat{f}$ & $0.38 \pm 0.01$ & $0.37 \pm 0.01$  & $-$\\
         $\chi^2/dof$& $1.3$ & $1.3$ & $2.5$\\
\hline \hline  & \multicolumn{3}{c}{Derived
   Observables}  \\
   $A_{1/2}^{p^*}[10^{-3}\text{GeV}^{-1/2}]$ & $ -9.7 \pm 4.1 $ & $-8.8
\pm 3.8$  & $-7.3 \pm 3.0$  \\
         $A_{3/2}^{p^*}[10^{-2}\text{GeV}^{-1/2}]$ & $2.3 \pm 1.1$ &
$2.1 \pm 1.0 $ & $2.5 \pm 0.8$ \\
         $\Gamma_{N^* \to \Lambda^*K}$ [MeV] & $22 \pm 7$ & $25\pm 7$ &
$30 \pm 8 $ \\
         $\frac{\Gamma_{N^* \to \Lambda^*K}}{\Gamma_{N^*}}$[$\%$] &
$12.9 \pm 3.9 $ &$ 14.8 \pm 4.5$  & $16.2 \pm 4.2 $ \\
\hline \hline
\end{tabular} \label{tab:nstar}
\end{table*}

\begin{figure*}[htbp]
\begin{center}
\makebox[0pt]{\includegraphics[scale=1.2]{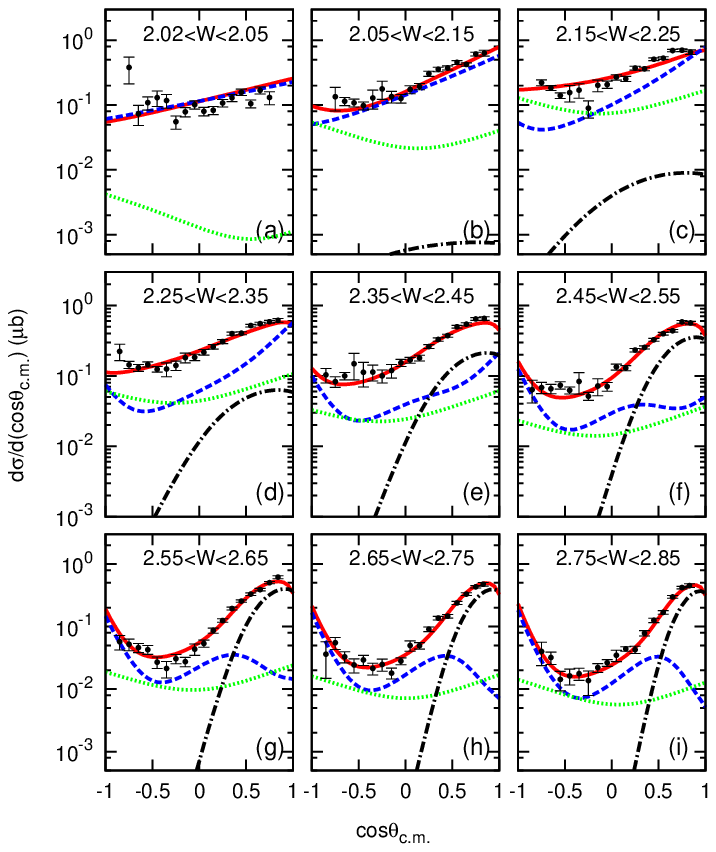}
\hspace{1cm}\includegraphics[scale=1.2]{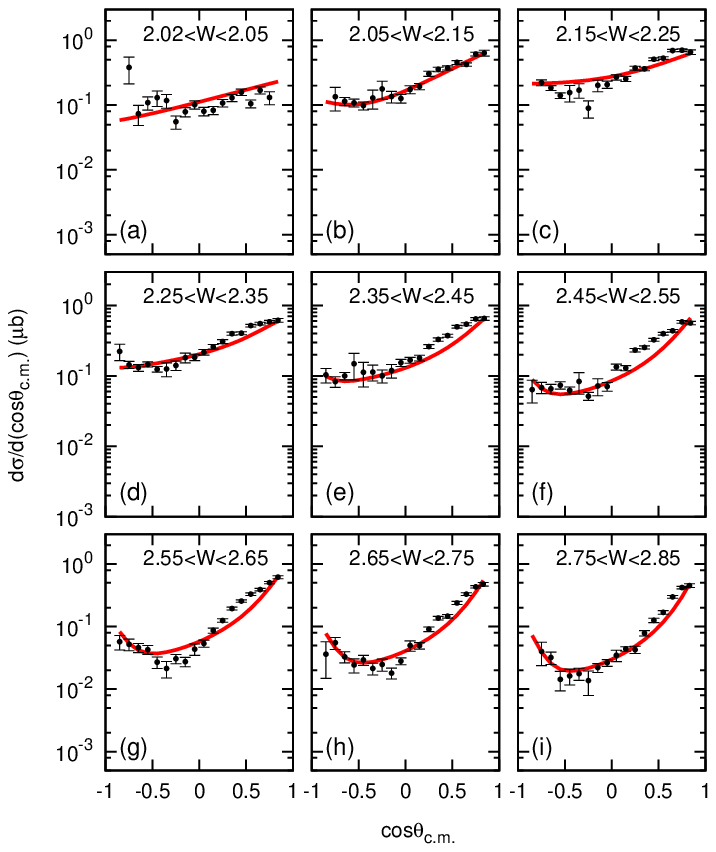}}
\vspace{-0.5cm} \caption{(Color online) Left: Model B $\gamma p \to
K^+ \Lambda(1520)$ differential cross sections as a function of
$\cos\theta_{\rm c.m.}$ compared with the CLAS
data~\cite{Moriya:2013hwg} for different $\gamma p$ invariant mass
intervals (indicated in the different panels in GeV units). Only
statistical errors are displayed. The blue-dashed and
black-dash-dotted curves stand for the contributions from the
effective Lagrangian approach background and Reggeon exchange
mechanism, respectively (see text for details). The green-dotted
lines show the contribution of the $N^*(2120)$ resonance term, while
the red-solid lines display the results obtained from the full
model. Right: Total results from our previous Fit II carried out in
Ref.~\cite{Xie:2013mua}), where Regge effects were not considered.}
\label{dcs-clas}
\end{center}
\end{figure*}

\begin{figure*}[htbp]
\begin{center}
\makebox[0pt]{\includegraphics[scale=0.8]{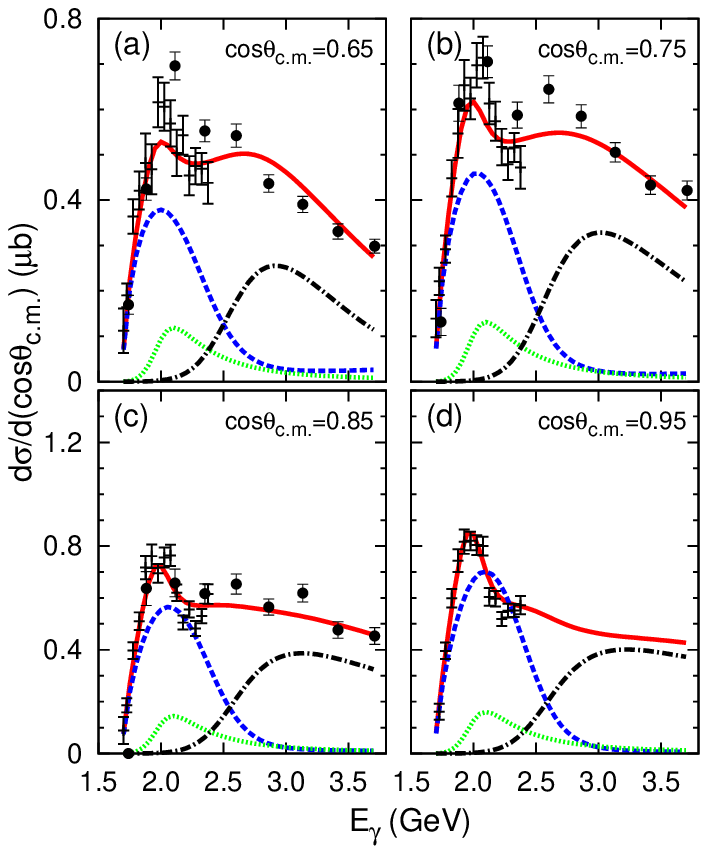} \hspace{0.5cm}
\includegraphics[scale=0.8]{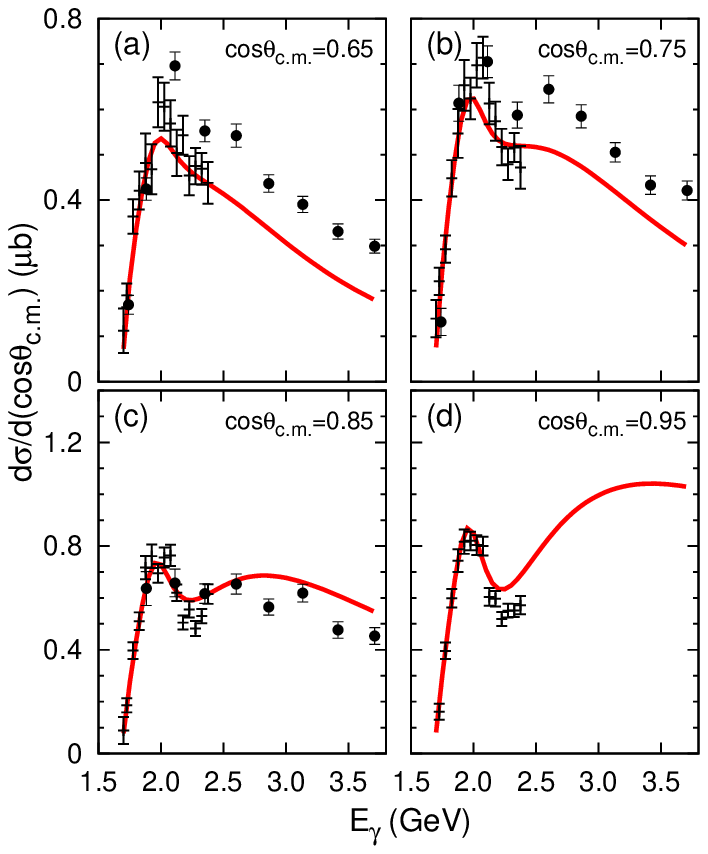}} \vspace{-0.5cm}
\caption{(Color online) Left: Model B $\gamma p \to K^+ \Lambda
(1520)$ differential cross section as a function of the LAB frame
photon energy for different c.m. $K^+$ polar angles. We also show
the experimental LEPS~\cite{leps2} (crosses) and
CLAS~\cite{Moriya:2013hwg} (black dots) data. Only statistical
errors are displayed. The blue-dashed and black-dash-dotted curves
stand for the contributions from the effective Lagrangian approach
background and Reggeon exchange mechanism, respectively (see text
for details). The green-dotted lines show the contribution of the
$N^*(2120)$ resonance term, while the red-solid lines display the
results obtained from the full model. Right: Total results from our
previous Fit II carried out in Ref.~\cite{Xie:2013mua}), where Regge
effects were not considered.} \label{dcs-leps}
\end{center}
\end{figure*}

We have performed a ten-parameter ($g_1$, $g_2$, $\Lambda_R$,
$ef_1$, $ef_2$, $M_{N^*}$, $\Gamma_{N^*}$, $t_0$, $\Delta t$ and
$\hat{f}$) $\chi^2-$fit to the  LEPS~\cite{leps2} and
CLAS~\cite{Moriya:2013hwg} measurements of
$d\sigma/d(\cos\theta_{\rm c.m.})$. There is a total of 216
available data (157 points from CLAS and another 59 ones from LEPS,
depicted in Figs.~\ref{dcs-clas} and \ref{dcs-leps}, respectively).
The systematical errors of the experimental data
($11.6\%$~\cite{Moriya:2013hwg} and $5.92\%$~\cite{leps2}, for CLAS
and LEPS, respectively) have been added in quadratures to the
statistical ones and taken into account in the  fits, as it was done
in Ref.~\cite{Xie:2013mua}. LEPS data lie in the $K^+$ forward angle
region and were taken below $E_\gamma$ = 2.4 GeV, while the recent
CLAS measurements span a much larger $K^+$ angular and photon energy
regions (nine intervals of the $\gamma p$ invariant mass from the
reaction threshold, 2.02 GeV, up to 2.85 GeV)~\footnote{To compute
the cross sections in each interval, we always use the corresponding
mean value of $W$, as in Ref.~\cite{Xie:2013mua}.}.

We have considered two different schemes to include  Regge effects
(models A and B), as discussed in Subsec.~\ref{sec:regge}. Best fit
results are listed in Table~\ref{tab:nstar}, where we also compile
the obtained parameters in our previous work (Fit II of
Ref.~\cite{Xie:2013mua}). For each fit, we also give the predicted
$N^*(2120)$ partial decay width $\Gamma_{N^* \to \Lambda^*K}$ (Eq.~(18)
of Ref.~\cite{Xie:2010yk}) and the resonance helicity amplitudes (Eqs.~(15) and (16)
of Ref.~\cite{Xie:2010yk}) for the positive-charge state.

A $\chi^2/dof$ around 1.3 is obtained for both model A and B fits.
This is significantly better than the best fit value obtained (2.5)
in our previous work of Ref.~\cite{Xie:2013mua}, where Regge effects
were not considered. We also see that the effective Lagrangian
approach parameters ($g_1$, $g_2$, $\Lambda_R$, $ef_1$, $ef_2$,
$M_{N^*}$, $\Gamma_{N^*}$), determined in the new fits carried out
in this work, turn out to be in good agreement with those obtained
in Ref.~\cite{Xie:2013mua}. Thus, the conclusions of that reference
still hold, in particular this new study gives further support to
the existence of the two-star $N^*(2120)$ resonance, and its
relevance in the CLAS \& LEPS $\gamma p \to K^+ \Lambda(1520)$ data.
On the other hand, the hybrid model parameters ($t_0$, $\Delta t$
and $\hat{f}$) turn out to be reasonable from what one would expect
by a direct inspection of the CLAS data ($t_0$, $\Delta t$) and
previous estimates~\cite{Grishina:2005cy,toki}.

The fits obtained here are of similar quality to the best ones
reported in Ref.~\cite{He:2014gga}, where in addition to the Regge
effects driven by kaon exchange in the $t-$channel, some sizable
Regge contributions induced by $\bar K^*$ exchanges are included as
well. However, as mentioned in the introduction, theoretically it is
difficult to accommodate a $\bar K^*$ mechanism contribution as
large as that claimed in Ref.~\cite{He:2014gga} (see sections 3.1
and 3.2 of this latter reference). On the other hand, a bunch of
$N^*$ resonances are included in the approach followed in
Ref.~\cite{He:2014gga}. Their couplings and masses are in most cases
fixed to the constituent quark model predictions of
Refs.~\cite{Capstick:1992uc,simonprd58} and a common width of 330
MeV is assumed for all of them. Among all of them, it turns out to
be the $N^*(2120)$, the state that provides the most important
contribution, which confirms previous
claims~\cite{Xie:2010yk,He:2012ud} . We have adopted a different
point of view and have used the accurate CLAS \& LEPS $\gamma p \to
K^+ \Lambda(1520)$ data not only to claim the existence of the
$N^*(2120)$ resonance, but also to establish some of its properties.
Thus, we find a much narrower state ($\Gamma_{N^*} \sim 170-175$
MeV) and complete different helicity amplitudes. Moreover, the
values used in Ref.~\cite{He:2014gga} ($A_{1/2}^{p^*}=36$ and
$A_{3/2}^{p^*}=-43$ in $[10^{-3}\text{GeV}^{-1/2}]$ units) are
incompatible both with
\begin{eqnarray}
A_{1/2}^{p^*}[10^{-3}\text{GeV}^{-1/2}] &=& 125 \pm 45 \\
A_{3/2}^{p^*}[10^{-2}\text{GeV}^{-1/2}] &=& 15 \pm 6\,,
\end{eqnarray}
given in Ref.~\cite{Anisovich:2011fc} and with previous
measurements~\cite{Awaji:1981zj}
\begin{eqnarray}
A_{1/2}^{p^*}[10^{-3}\text{GeV}^{-1/2}] &=& -20 \pm 8 \label{eq:a12} \\
A_{3/2}^{p^*}[10^{-2}\text{GeV}^{-1/2}] &=&  1.7 \pm 1.1  \label{eq:a32}
\end{eqnarray}
quoted in the 2008 PDG edition~\cite{Amsler:2008zzb}, that in turn
are in quite good agreement with our predictions in
Table~\ref{tab:nstar}. Having improved the quality of our fit,
achieving now an accurate description of the CLAS data for all
angles and invariant mass windows (see below), our results give an
important support to the measurements of Ref.~\cite{Awaji:1981zj},
which do not seem entirely consistent with those reported in
Ref.~\cite{Anisovich:2011fc}. Given the two stars status (evidence
of existence is only fair) granted to the $N^*(2120)$ resonance in
the multichannel partial wave analysis of pion and photo-induced
reactions off protons carried out in Ref.~\cite{Anisovich:2011fc},
the discrepancy with our predicted helicity amplitudes should not be
used to rule out our fits, but rather one should interpret our
results as further constrains on these elusive observables. Note
that the helicity amplitudes given in Eqs.~(\ref{eq:a12}) and
(\ref{eq:a32}) were also used in Ref.~\cite{Nam:2012ui}, where the
$ep \to e K^+ \Lambda(1520)$ CLAS data of Ref.~\cite{Barrow:2001ds}
was successfully described.

In addition, there is a disturbing feature in the fits presented in
\cite{He:2014gga}. There, it is found $t_0\sim 3$ GeV$^2$, though
with a large error, while we obtain values in the range 0.7--0.9
GeV$^2$. A value of $t_0$ as high as $3$ GeV$^2$ necessarily changes
the meaning of the interpolating function ${\cal R}_t$ in
Eq.~(\ref{eq:rt}), since it will not effectively filter now forward
angles. This is easily understood if one realizes that for $W=2.4$
GeV, $|t|$ remains below 2.5 GeV$^2$ for all possible $K^+$ c.m.
angles, and for the highest invariant mass $W=2.8$ GeV, the bound
$t=-3$ GeV$^2$ is reached for $\cos\theta_{\rm c.m.}=-0.3$. Thus in
the scheme employed in \cite{He:2014gga}, the transition function
${\cal R}_t$ effectively modifies the predictions of the effective
Lagrangian approach allowing for some Regge effects for large
scattering angles, which seems quite doubtful. Probably, this
dis-function of the physical meaning of ${\cal R}_t$ could be a
consequence of the unnecessary complexity of the scheme used in
Ref.~\cite{He:2014gga} with various $N^*$ contributions and the
inclusion of $\bar K^*$ driven effects,  with parameters in some
cases fixed to values with little theoretical/experimental support.
Nevertheless, it should be acknowledged that the work of
Ref.~\cite{He:2014gga} is pioneer in exploring the possible
existence of Regge effects in the CLAS data.

The differential $d\sigma/d(\cos\theta_{\rm c.m.})$ distributions
calculated with the model B best-fit parameters are shown in
Figs.~\ref{dcs-clas} and ~\ref{dcs-leps} as a function of
$\cos\theta_{\rm c.m.}$ and for various $\gamma p$ invariant mass
intervals. Model A results are totally similar and for brevity, they
will not be discussed any further. Only statistical errors are
displayed in these two figures and the contributions from different
mechanisms are shown separately. Thus, we split the full result into
three main contributions: effective Lagrangian approach background,
Reggeon exchange and resonance $N^*(2120)$. The first one
corresponds to the $t$-channel $\bar K$ exchange, nucleon pole,
contact and $u$-channel $\Lambda(1115)$ hyperon pole terms of
Eqs.~(\ref{eq:at})--(\ref{eq:ac}) and (\ref{eq:au}), but evaluated
with the modified form-factor $f_c(1-{\cal R})$  instead of $f_c$,
as discussed in Eq.~(\ref{eq:smooth}). (Note that $f_c$ appears
neither in the $\Lambda(1115)$ nor in the resonance $N^*(2120)$
mechanisms because both of them  are gauge invariant by themselves).
The Reggeon contribution is calculated from the $f_c$ terms of the
$\bar K$ exchange,  nucleon pole and  contact terms of
Eqs.~(\ref{eq:at})--(\ref{eq:ac}) and (\ref{eq:au}), but now
evaluated with the generalized Regge form-factor  ${\cal F}^{\,\rm
Regg}_{B}{\hat f} {\cal R}$.

In the left panels of the first of these two figures, we show our
predictions and the data of the CLAS
collaboration~\cite{Moriya:2013hwg}. In the right panels and for
comparison purposes, we display the final results from our previous
Fit II carried out in Ref.~\cite{Xie:2013mua}, where Regge effects
were not considered. We find an overall good description of the data
for the whole range of measured $\gamma p$ invariant masses  and it
is significantly better than that exhibited in the right panels.  We
see that the Regge improved model provides now an excellent
description of the CLAS data for values of $\cos\theta_{\rm c.m.}$
above 0.5, and high energies, $W\ge 2.3$ GeV, as expected. On the
other hand, by construction Regge contributions effectively
disappear at low invariant masses $W<2.3$ GeV and backward $K^+$
angles. Thus, we recover for this latter kinematics the effective
Lagrangian approach, including resonance $N^*(2120)$ and hyperon
$\Lambda(1115)$ contributions, which successfully described the data
in this region~\cite{Xie:2013mua}.

In the left panels of Fig.~\ref{dcs-leps}, the differential cross
section deduced from the results of the model B fit, as a function
of the LAB frame photon energy and for different forward c.m. $K^+$
angles, is shown and compared both to LEPS~\cite{leps2} and
CLAS~\cite{Moriya:2013hwg} datasets. In the right panels and for the
sake of clarity, we display the final results from our previous Fit
II carried out in Ref.~\cite{Xie:2013mua}, where Regge effects were
not considered. We see the description of LEPS data is almost not
affected by the Regge contributions, and the bump structure in the
differential cross section at forward $K^+$ angles is fairly well
described thanks to the significant contribution from the $N^*$
resonance in the $s-$channel, as pointed out in
Ref.~\cite{Xie:2010yk,Xie:2013mua}. However, the inclusion of Regge
effects significantly improves the description of the CLAS
data~\footnote{The CLAS cross sections shown in the figure were
obtained from the appropriate CLAS measurements displayed in
Fig.~\ref{dcs-clas}, relating $W$ to the LAB photon energy.}, as one
would expect from the discussion of the results of
Fig.~\ref{dcs-clas}. Moreover, the hybrid model presented in this
work provides a better energy behavior for the forward cross section
at energies higher than those explored by the CLAS data (see the two
(d) panels in Fig.~\ref{dcs-leps}).

\begin{figure}[htbp]
\begin{center}
\includegraphics[scale=0.8]{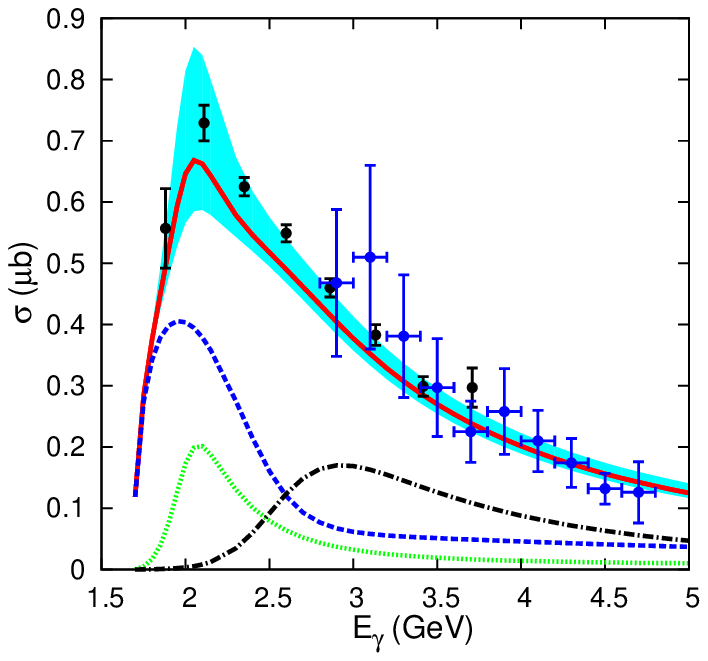} \vspace{-0.5cm}
\caption{(Color online) Total $\gamma p \to K^+ \Lambda^*$ cross
section as a function of the photon energy. Black filled circles and
blue open circles stand for CLAS~\cite{Moriya:2013hwg} and
LAMP2~\cite{Barber:1980zv} data, respectively. LAMP2 cross sections
have been scaled down by a factor 0.6. Results from model B are also
shown: The blue-dashed and black-dash-dotted curves stand for the
contributions from the effective Lagrangian approach background and
Reggeon exchange mechanism, respectively (see text for details). The
green-dotted lines show the contribution of the $N^*(2120)$
resonance term, while the red-solid lines display the results
obtained from the full model. The shaded region accounts for the
68\% CL band inherited from the Gaussian correlated statistical
errors of the parameters. } \label{Fig:tcs}
\end{center}
\end{figure}

Fig.~\ref{Fig:tcs} shows the $\Lambda(1520)$ total photoproduction
cross section as a function of the photon energy. Despite the
overall normalization of the CLAS~\footnote{We display extrapolated
total cross sections, from data summed over the useful acceptance of
the detector, to 4$\pi$ (red points in Fig. 11 of
Ref.~\cite{Moriya:2013hwg}).} measurements~\cite{Moriya:2013hwg}  is
in rather strong disagreement with the data from
LAMP2~\cite{Barber:1980zv}, the photon energy dependence of both
data sets seems compatible above 2.3 or 2.4 GeV. This can be
appreciated in Fig.~\ref{Fig:tcs}, where the LAMP2 cross sections
have been scaled down by a factor 0.6. This agreement might give
some support to the idea of finding Regge signatures in the CLAS
data. Results from model B are also shown, which turn out to provide
a good description of both sets of data. We should, however, prevent
the reader about the {\it ad hoc} modification of the normalization
of the old LAMP2 cross sections~\footnote{The low energy SAPHIR
data~\cite{Wieland:2011zz} is in even in a stronger disagreement
with the data from LAMP2, with the CLAS results lying almost exactly
between these two measurements~\cite{Moriya:2013hwg}.}.
Nevertheless, it is reassuring that the hybrid model presented in
this work, including Regge effects, is able to predict the photon
energy dependence of the LAMP2 data at energies well above than
those explored by the CLAS data.

\section{conclusions} \label{sec:conclusions}

We have presented some evidences of Regge signatures in the CLAS
data at forward angles, despite the energies involved in that
experiment are only moderately high. This is not entirely
surprising, because above $E_\gamma > 2.3-2.4$ GeV, and up to an
overall normalization, the CLAS $\Lambda(1520)$ total cross section
dependence on the photon energy matches that inferred from the LAMP2
data, which extends up to 5 GeV, in a region where the Regge
behavior is expected to be visible (see Fig.~\ref{Fig:tcs}). Indeed,
we find a significant improvement on the description of the CLAS
high energy forward cross sections, when the effective Lagrangian
approach of Ref.~\cite{Xie:2013mua} is supplemented with some string
quark--gluon mechanism contributions determined by a kaon
trajectory. Now, there are no visible systematic discrepancies
between the hybrid approach predictions and the data. Thus, we
confirm the findings of the recent work of Ref.~\cite{He:2014gga} on
the importance of the Regge effects in achieving an accurate
description of the CLAS forward angular distributions.

We do not need to include any contribution from a $\bar K^*$
trajectory, in accordance to the analysis of the LAMP2 data carried
out in Refs.~\cite{toki,Nam:2010au}. This is re-assuring since the
$t-$channel $\bar K^*$ contribution should be quite small, almost
negligible, in sharp contrast with previous works
~\cite{Nam:2005uq,titovprc7274,sibiepja31,He:2014gga}, where a large
$g_{K^*N\Lambda^*}$ coupling was assumed. Such big values for this
coupling are ruled out by unitarized chiral
models~\cite{toki,Hyodo:2006uw,Gamermann:2011mq}, that predict
values for $g_{K^*N\Lambda^*}$ around a factor 10 (20) smaller than
for instance those used in
Refs.~\cite{Nam:2005uq,titovprc7274,He:2014gga}, and by measurements
of the photon-beam asymmetry, as discussed in
Ref.~\cite{Nam:2010au}.

We have designed a gauge invariant hybrid model which smoothly
interpolates from the hadron effective Lagrangian
approach~\cite{Xie:2013mua}, at energies close to threshold, to a
model that incorporates quark-gluon string reaction mechanism
contributions at high energies and forward $K^+$ scattering angles.
We find an accurate description of both CLAS and LEPS data. The
latter set of low energy cross sections is not affected by the
inclusion of Regge effects. The bump structure observed at forward
$K^+$ angles in these data is well described thanks to the
significant contribution from the two-star $J^P=3/2^-$ $N^*(2120)$
resonance in the $s$-channel, which existence gets a strong support
from this improved analysis that is now fully consistent with the
accurate CLAS data. Thus, this associated strangeness production
reaction becomes an excellent tool to determine the properties of
this resonance (helicity amplitudes determined by the couplings
$ef_1$ and $ef_2$ or the strength of the $K\Lambda^*N^*$ vertex). In
what respects to the CLAS data, Regge effects play a crucial role at
forward angles for energies above 2.35 GeV, as commented before,
while the backward angle data highlight the importance of the
$u$-channel $\Lambda(1115)$ hyperon pole term. This latter fact can
be used to constrain the radiative $\Lambda^*\to \Lambda \gamma$
decay, as it was firstly emphasized in Ref.~\cite{Xie:2013mua}.

The $t$-range explored by the CLAS data is not large enough to fully
restrict the Regge form-factor, which is the major difference among
the two models (A and B) introduced in this work.  Though, in the
region of negative $t$, the Reggeized propagator in
Eq.~(\ref{eq:nam}) exhibits a factorial growth, which is in
principle not acceptable, the limited range of momentum transfers
accessible in the data does not see this unwanted behaviour. This is
the same reason why the Gaussian cutoff parameter $a$ in
Eq.~(\ref{eq:agauss}) is not further constrained. Unfortunately, the
existing large  discrepancies among CLAS and LAMP2 data sets
prevents the inclusion of this latter experiment in the analysis
carried out in this work. This constitutes an open problem, that
might require new dedicated experiments.

\section*{Acknowledgments}

This work was partly supported by DGI and FEDER funds, under
Contract No. FIS2011-28853-C02-01 and FIS2011-28853-C02-02, the
Spanish Ingenio-Consolider 2010 Program CPAN (CSD2007-00042),
Generalitat Valenciana under Contract No. PROMETEO/2009/0090, and by
the National Natural Science Foundation of China under Grant No.
11105126. We acknowledge the support of the European
Community-Research Infrastructure Integrating Activity Study of
Strongly Interacting Matter (HadronPhysics3; Grant Agreement No.
283286) under the Seventh Framework Programme of the E.U. The work
was supported in part by DFG (SFB/TR 16, Subnuclear Structure of
Matter.

\end{document}